\pdfoutput=1
\documentclass[11pt,a4paper]{article}

\usepackage[T1]{fontenc}
\usepackage[utf8]{inputenc}
\usepackage{lmodern}
\usepackage{amsmath,amssymb}
\usepackage[margin=2.5cm]{geometry}
\usepackage{microtype}
\usepackage{graphicx}
\usepackage{authblk}
\usepackage{booktabs}
\usepackage{xcolor}
\usepackage{tikz}
\usetikzlibrary{positioning,arrows.meta,fit,backgrounds,calc}
\usepackage[colorlinks=true,linkcolor=blue!60!black,citecolor=blue!60!black,urlcolor=blue!60!black]{hyperref}

\newcommand{\red}[1]{#1}
\newcommand{\green}[1]{#1}

\title{Computing the Wave: Bridging Gravitational Wave and
High-Energy-Physics software (with modular ROOT extensions)}

\author[a,b]{Marco Meyer-Conde\thanks{Corresponding author: \href{mailto:marco@tcu.ac.jp}{marco@tcu.ac.jp}}}
\author[a]{Hirotaka Takahashi}
\author[c]{Nobuyuki Kanda}
\author[d]{Ken-ichi Oohara}
\author[e]{Kazuki Sakai}
\author[ ]{on behalf of the LIGO Scientific, Virgo and KAGRA Collaborations}

\affil[a]{Tokyo City University, 1-28-1 Tamazutsumi, Setagaya-ku, Tokyo, Japan}
\affil[b]{Department of Physics, University of Illinois at Urbana-Champaign, Urbana, Illinois, USA}
\affil[c]{Nambu Yoichiro Institute of Theoretical and Experimental Physics (NITEP), Osaka Metropolitan University, Osaka, Japan}
\affil[d]{Graduate School of Science and Technology, Niigata University, Niigata, Japan}
\affil[e]{Department of Electronic Control Engineering, National Institute of Technology (NIT), Nagaoka College, Nagaoka, Niigata, Japan}

\date{Contribution to the 23rd International Workshop on Advanced Computing and
Analysis Techniques in Physics Research (ACAT2025),\\
8--12 September 2025, Hamburg, Germany. To appear in Proceedings of Science.}

\begin{document}
\maketitle

\begin{abstract}
High-Energy Physics (HEP) and Gravitational Wave (GW) astronomy
pursue different scientific goals, yet they face strikingly similar computing
challenges and can enrich one another through shared software. Building on the
prototype presented at ACAT 2024, we report on the consolidation of that work
into a single, modular extension of the CERN ROOT data-analysis framework,
\green{freely distributed} under \texttt{RIOPlus}. The library is now organised as a
set of optional, dynamically loadable extensions covering configuration and
file-format parsing (including the \green{Gravitational Wave Frame (GWF)} format and HDF5), \green{remote and federated data access} (OSDF/Pelican, S3, Kafka) with token-based credentials,
an enhanced dynamic-display manager, an advanced signal-processing layer built on
the KFR library, and a complex-capable tensor layer backed by LibTorch with ONNX
interoperability. We summarise the design of each extension, \green{outline the intended}
application to transient-noise (\emph{glitch}) detection and classification, and
update the FFT performance benchmark comparing KFR with FFTW. We close with
\green{the} end-to-end ROOT-centred workflow \green{these extensions enable, its current status,} and an outlook toward
column-wise RNTuple storage.
\end{abstract}

\section{Introduction}
\label{sec:intro}

The timely detection of compact-binary mergers is of central importance to both astrophysics and nuclear physics. The first binary-neutron-star merger,
GW170817 together with its electromagnetic counterpart GRB170817A~\cite{gw170817}, opened the era of multi-messenger astronomy and demonstrated the value of low-latency alerts. Over the last years the main gravitational wave (GW) collaborations have consolidated their computing efforts into a common infrastructure, the International Gravitational-Wave Observatory Network (IGWN)~\cite{igwn}, jointly operated by the LIGO~\cite{aligo}, Virgo~\cite{avirgo} and KAGRA~\cite{kagra} collaborations.

\green{IGWN and the large HEP experiments (e.g.\ ATLAS and CMS at the LHC)} share many computing challenges --- federated data distribution, heterogeneous pipelines, the integration of modern machine learning --- even though their requirements differ in emphasis. High-energy physics (HEP) is dominated by the need to process very large data volumes, whereas the GW community is driven primarily by latency: science data streams are comparatively small (a few calibrated channels), but they must be transported and analysed in near real time to enable rapid sky localisation and follow-up. The goal of the present work, continuing the prototype reported at ACAT~2024~\cite{acat2024}, is to get the best of both worlds by bringing the maturity and performance of HEP analysis software --- in particular the CERN ROOT framework~\cite{root} --- to GW data analysis, while keeping the result lightweight, optional and easy to adopt.

ROOT has already been used within the GW community, mostly by researchers with a
HEP background; a concrete and successful example is the Omicron
software~\cite{omicron}, which performs the $Q$-transform on top of ROOT\green{;
the cWB pipeline~\cite{cwb} likewise uses ROOT alongside the LIGO Algorithm
Library Suite (LALSuite)~\cite{lalsuite}, the C package that underpins most GW
pipelines}. Rather than replacing existing
tools, our aim is to extend ROOT so that it covers the features a GW workflow
requires. From discussions with pipeline development researchers we distilled six practical requirements \green{useful for both communities}:

\begin{enumerate}
\item \emph{common software distribution} for Python, C/C++ and Fortran;
\item a coherent \emph{ecosystem} and packaging story;
\item \green{\emph{remote data access}}: configuration files, remote storage and the
      GWF/HDF5 file formats;
\item \emph{enhanced dynamic display} with generic precomputed layout;
\item \emph{advanced signal processing}; and
\item \emph{complex-valued tensors} with low-maintenance machine-learning support.
\end{enumerate}

\green{The immediate motivation is transient-noise characterisation in KAGRA:
glitch classification for O3GK was done offline with unsupervised
learning~\cite{sakai2022,sakai2024,oshino2025}, and linking glitch families to
their instrumental origin needed a separate veto study~\cite{hveto}. The steps
involved --- reading distributed data, time-frequency analysis, inference with
Python-trained models --- are spread across tools in different languages and
data containers. This proceeding is limited to the software addressing this;
scientific results obtained with it are left to dedicated publications.}

\section{A modular ROOT extension: \texttt{RIOPlus}}
\label{sec:riioplus}

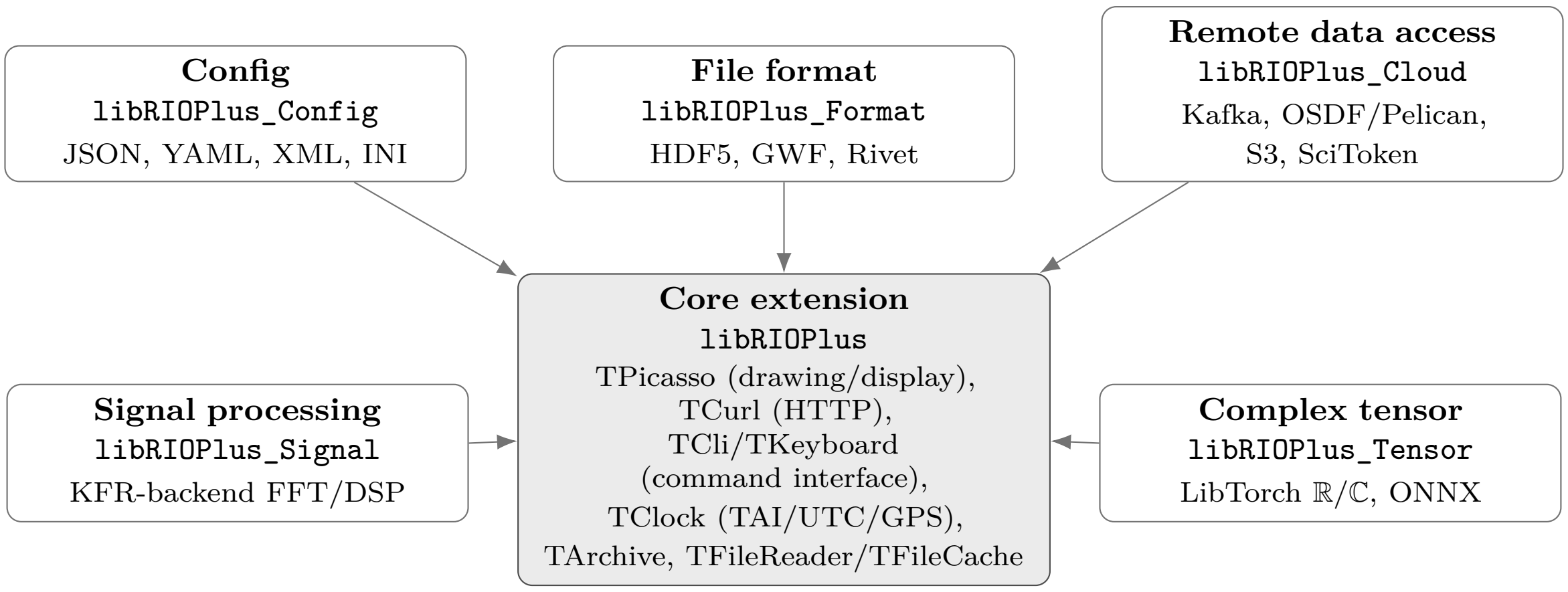
\begin{figure}[htbp]
\centering
\begin{tikzpicture}[
   font=\footnotesize,
   core/.style={rectangle, rounded corners, draw=black!70, fill=black!8,
                text width=5cm, align=center, minimum height=1.2cm},
   ext/.style={rectangle, rounded corners, draw=black!55, fill=white,
               text width=4.3cm, align=center, minimum height=1.25cm},
   line/.style={-{Latex[length=2mm]}, draw=black!55}
]

\node[core] (core) {\textbf{Core extension}\\
\texttt{libRIOPlus}\\[1pt]
{\scriptsize
TPicasso (drawing/display),\\TCurl (HTTP),\\
TCli/TKeyboard (command interface),\\
TClock (TAI/UTC/GPS), TArchive, TFileReader/TFileCache
}};

\node[ext, above left=0.9cm and 0.5cm of core] (config)
{\textbf{Config}\\
\texttt{libRIOPlus\_Config}\\[1pt]
{\scriptsize JSON, YAML, XML, INI}};

\node[ext, above=0.9cm of core] (format)
{\textbf{File format}\\
\texttt{libRIOPlus\_Format}\\[1pt]
{\scriptsize HDF5, GWF, Rivet}};

\node[ext, above right=0.9cm and 0.5cm of core] (cloud)
{\textbf{Remote data access}\\
\texttt{libRIOPlus\_Cloud}\\[1pt]
{\scriptsize Kafka, OSDF/Pelican, S3, SciToken}};

\node[ext, below=-2cm of core, xshift=-5.4cm] (signal)
{\textbf{Signal processing}\\
\texttt{libRIOPlus\_Signal}\\[1pt]
{\scriptsize KFR-backend FFT/DSP}};

\node[ext, below=-2cm of core, xshift=5.4cm] (tensor)
{\textbf{Complex tensor}\\
\texttt{libRIOPlus\_Tensor}\\[1pt]
{\scriptsize LibTorch $\mathbb{R}/\mathbb{C}$, ONNX}};

\foreach \x in {config,format,cloud,signal,tensor}
    \draw[line] (\x) -- (core);
\end{tikzpicture}

\caption{Modular structure of the \texttt{RIOPlus} ROOT extension. A core
library provides general-purpose features; each extension is a separately
loadable shared object addressing one of the requirements of
section~\ref{sec:intro}.}
\label{fig:architecture}
\end{figure}

Following ACAT~2024 the prototype libraries were consolidated and made available
at \url{https://git.ligo.org/kagra/libraries-addons/root/root-plus}. The code is now
organised around a core library, \texttt{libRIOPlus}, complemented by a set of optional extensions that the user loads only when needed
(figure~\ref{fig:architecture}). Each extension is an independent shared object (\texttt{.so}), so a deployment pays only for the features it uses and the additional dependencies they entail. All extensions add high-level helpers \green{with no runtime overhead} on top of standard ROOT classes, remaining fully compatible with the rest of the framework.

\subsection{Core extension}
The core library bundles general-purpose helpers used across all
workflows: a drawing manager (\texttt{TPicasso}), a libcurl HTTP client
(\texttt{TCurl}), a command-line and keyboard helper (\texttt{TCli}/\texttt{TKeyboard}),
a clock handling the TAI, UTC and GPS time systems used in GW data
(\texttt{TClock}), and an on-the-fly compressed-archive helper (\texttt{TArchive}).

\subsection{Configuration and file-format extensions}
The configuration extension \green{(\texttt{libRIOPlus\_Config})} exposes a single adapter interface
(\texttt{TConfig}, with concrete back-ends such as
\texttt{TConfigJSON}) so that JSON, YAML, XML and INI files can be read and written interchangeably. The file-format extension \green{(\texttt{libRIOPlus\_Format})} adds support for reading and writing HDF5 (the de-facto format for GW open-science data) and the \green{Gravitational Wave Frame (GWF) format~\cite{frameformat}}, the standard low-latency strain format in which one calibrated second of data is stored per frame. Rivet files are also supported, easing interoperability with HEP tooling.

\subsection{\green{Remote data access} and credentials extension}
Modern GW analysis is increasingly federated. \green{This} extension integrates
several remote-storage \green{and streaming} back-ends behind ROOT-like directory and file abstractions:
the Open Science Data Federation (OSDF)~\cite{osdf}, accessed over HTTP through the
Pelican protocol (\texttt{TDirectoryOSDF}); S3-compatible object storage based on
the open-source MinIO client (\texttt{TDirectoryS3}); and Apache Kafka~\cite{kafka} producers
and consumers (\texttt{TKafka}), with message packing for long payloads. \green{Kafka being payload-agnostic, the chunked GWF messages of the IGWN
low-latency data distribution (IGWN-LLDD) could use the same packing layer;
this is not exercised here.} Access to
distributed resources worldwide is handled through JSON Web Token and SciToken
credentials (\texttt{TJWToken}/\texttt{TSciToken}), used together with the
\texttt{TCurl} interface.

\subsection{Enhanced dynamic display}
GW publications and dashboards rely on a recurring set of plots (running
time-series, ratio plots, confusion matrices, skymap projections). The display helpers, built on the ROOT \texttt{TExec} mechanism, handle in-sync axis, tick and label positioning from margin-and-padding information, support shared axes, and keep a fixed or dynamic size on demand so that a canvas remains portable when resized.

\subsection{Advanced signal-processing extension}
\label{sec:signal}
The signal-processing extension is the part most specific to GW analysis. It wraps the KFR library~\cite{kfr}, a modern, actively maintained C++ DSP framework released under the GPL, behind a \texttt{TKFR} interface that is compatible with ROOT's \texttt{TVirtualFFT} (while remaining optional). KFR provides $N$-dimensional DFT/FFT together with a rich set of signal-processing primitives: amplitude- and power-spectral densities, the Hilbert transform, FIR and IIR filtering, windowing, decimation, and waveform generation, and analog/digital filter design (Butterworth, Chebyshev, elliptic) through \texttt{TComplexFilter}, \texttt{TBiquad} and \texttt{TPoleZeroMap}. They are exposed through \texttt{TTimeSeries} and \texttt{TFrequencySeries} (both derived from the \texttt{TTensor<Element>} introduced below), covering use cases from spectral estimation to matched filtering. The library is portable across architectures and exploits SIMD acceleration through SSE/SSE2/AVX instructions on x86\texttt{-}64 and NEON on ARM64.

\begin{figure}[htbp]
    \centering
    \includegraphics[width=1\textwidth]{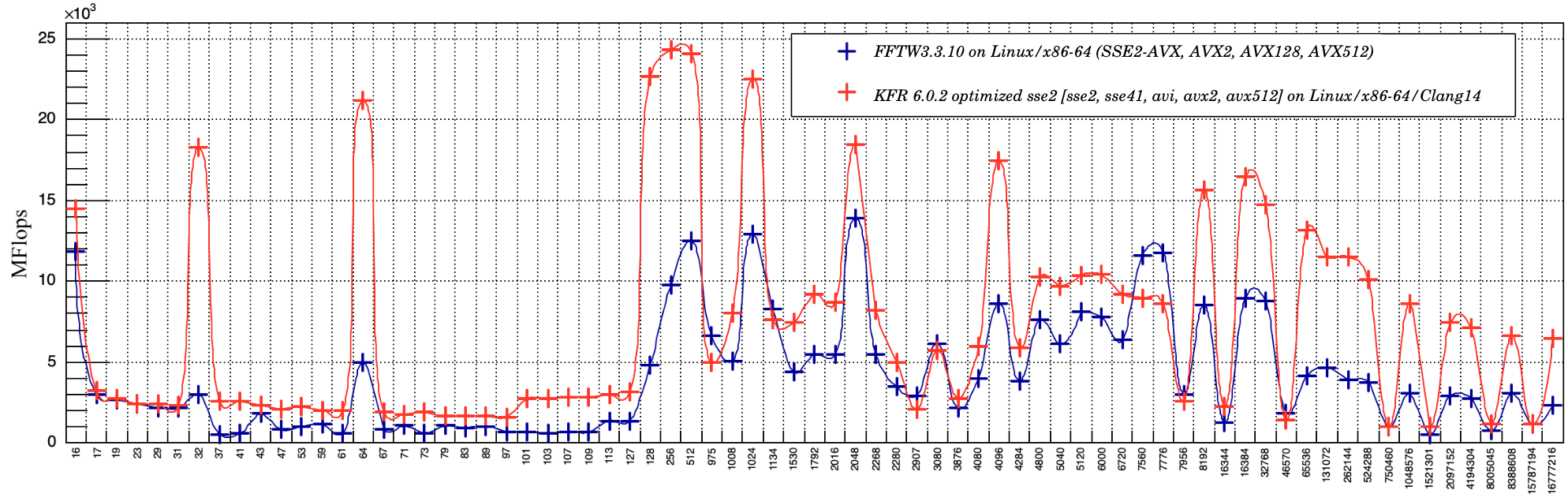}
    \caption{MFlops distributions (the higher, the better) vs FFT Size from a benchmark test performed on a local cluster in Osaka, comparing complex forward transform (in-place) between FFTW3.3.10 and KFR6.0.2}
    \label{fig:kfrlib}
\end{figure}

\red{Figure~\ref{fig:kfrlib} shows a benchmark we developed to validate this
choice of backend (the \texttt{TKFR} wrapping overhead being negligible),
comparing an in-place complex forward transform between
FFTW~3.3.10~\cite{fftw} and KFR~6.0.2. The benchmark application and
instructions are publicly available at
\url{https://git.ligo.org/kagra/software/fft-benchmark}}

\subsection{Complex tensors and machine learning}
\label{sec:tensor}
The tensor extension provides a \texttt{TTensor<Element>} class for tensor and matrix computations backed by the LibTorch C++ API~\cite{libtorch}, sharing the same
backend as PyTorch and therefore enabling transparent GPU acceleration.
Following ROOT conventions (e.g.\ \texttt{TH1F}/\texttt{TH1D}), the precision notation is extended to four numerical types: real single/double (\texttt{F}/\texttt{D}) and complex single/double (\texttt{C}/\texttt{Z}). Tensors are created from existing ROOT buffers without copying (\texttt{torch::from\_blob}). Trained models exported in the Open Neural Network Exchange (ONNX) format~\cite{onnx} can be loaded through a \texttt{TModel} wrapper (an ONNX Runtime session) and applied directly to \texttt{TTensor} data, allowing recent models to be used and shared across Python and C/C++ applications without bespoke maintenance --- updating LibTorch through Conda is sufficient. This complements\green{, without additional maintenance burden,} the machine-learning capabilities already available in ROOT through TMVA/SOFIE~\cite{sofie}, which remains an attractive option for reproducibility.
\green{The extension targets deployment, not training: applying a
Python-trained model, in the same process, to complex-valued data held in ROOT
containers. A MNIST~\cite{MNIST} example is shipped with the library.}

\section{Application: transient-noise detection and classification}
\label{sec:glitch}

\red{A representative use case that exercises several extensions at once is the
handling of \emph{glitches} --- short instrumental or environmental transients
that mimic or mask astrophysical signals. Detecting them starts from a
power-excess search\green{, such as the $Q$-transform performed by Omicron~\cite{omicron} on top of ROOT. The primitives of section~\ref{sec:signal} are its building blocks, but reproducing Omicron is not the purpose of this work and no such comparison is presented here}.}

\red{Once candidates are extracted, classifying their morphology helps diagnose
detector behaviour~\cite{hveto}. The publicly available Gravity Spy
catalogue~\cite{gravityspy} lists tens of recurring glitch classes
identified since the first observing
run~\cite{sakai2022,sakai2024,oshino2025}.}
Using the tensor extension (section~\ref{sec:tensor}), we explore a workflow in
which an unsupervised embedding --- Uniform Manifold Approximation and Projection
(UMAP) --- is first used to find clusters in the space of glitch spectrograms for
pre-tagging, after which a convolutional neural network performs the final
classification\green{, the architecture used offline in~\cite{sakai2022,oshino2025}, with the trained models exported to ONNX}. Performing this online would flag recurrent system failures (e.g.\ cooling or actuators) as they happen.

\section{Distribution: packaging and containers}
\label{sec:dist}

Conda has become a
de facto standard for software deployment within the GW community, and we use
it as a package manager to distribute pre-compiled binaries.

Two community efforts are relevant: a lightweight \href{https://github.com/hep-forge}{hep-forge} conda channel with minimal publication requirements (Linux, AMD64, and optional ARM64), and a broader \href{https://github.com/hep-packaging-coordination}{HEP packaging-coordination} initiative targeting multiple architectures and operating systems on conda-forge, both independent of the legacy \texttt{hepforge.org} service.
\green{Generic conda binaries target a baseline instruction set, but KFR (like
LibTorch and ONNX Runtime) embeds several architecture-specific code paths
selected at run time, so the performance-critical layer is not penalised.}

\red{Reference documentation and tutorials are in preparation for the first
tagged release, and external contributors are welcome through the LIGO
GitLab instance.}

\section{\green{Workflow, status and outlook}}
\label{sec:workflow}

\begin{figure}[htbp]
\centering
\begin{tikzpicture}[
  font=\footnotesize,
  node distance=4mm,
  box/.style={rectangle, rounded corners, draw=black!60, fill=black!6,
              align=center, minimum height=0.95cm, text width=2.4cm},
  io/.style={rectangle, draw=black!60, fill=white, align=center,
             minimum height=0.95cm, text width=1.9cm},
  ar/.style={-{Latex[length=2mm]}, draw=black!60}
]
\node[io] (gwf) {GWF frames\\ \scriptsize (LVK)};
\node[box, above=of gwf] (ingest) {Remote ingest\\ \scriptsize TKafka, OSDF, S3 native retrieval,\\ with TSciToken};
\node[box, right=of ingest] (store) {ROOT store\\ \scriptsize TTree $\rightarrow$ RNTuple};
\node[box, right=of store] (dsp) {Signal proc.\\ \scriptsize KFR / TKFR\\ Time/Frequency\\ Series};
\node[box, right=of dsp] (ml) {ML inference\\ \scriptsize LibTorch / ONNX\\ (CPU/GPU)};
\node[io, below=8mm of ml] (out) {End-user\\ analysis};

\draw[ar] (gwf) -- (ingest);
\draw[ar] (ingest) -- (store);
\draw[ar] (store) -- (dsp);
\draw[ar] (dsp) -- (ml);
\draw[ar] (ml) -- (out);
\end{tikzpicture}
\caption{ROOT-centred GW analysis workflow built from the
\texttt{RIOPlus} extensions, from low-latency frame ingestion to machine-learning
inference and end-user analysis.}
\label{fig:workflow}
\end{figure}
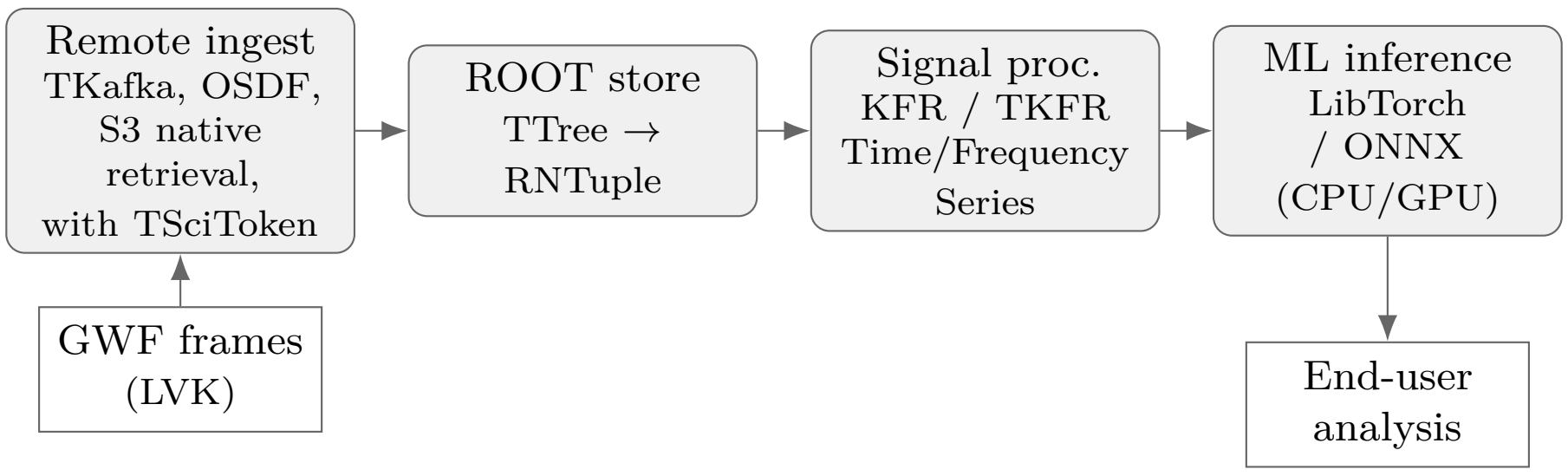

\green{The extensions compose into an end-to-end, ROOT-centred workflow for GW analysis (figure~\ref{fig:workflow}):} low-latency GWF frames are ingested and stored into \texttt{TTree}s as one-second \texttt{std::vector<double>} frames, distributed through the \green{remote-data} extension (\texttt{TDirectory}/\texttt{TFile} over OSDF or S3, with Kafka transport and SciToken credentials), processed by the KFR-backed signal-processing layer, and fed to machine-learning models via LibTorch/ONNX with GPU acceleration \green{already available through LibTorch}.

\red{The core extension including CLI, drawing, and clock helpers, together with the configuration and archive utilities and the \green{remote-data} distribution layer (\texttt{TCloud}/\texttt{TBucket}, \texttt{TDirectory} and \texttt{TFile} over OSDF/S3, \texttt{TSciToken}, and \texttt{TKafka}) has been completed, while the KFR-based \texttt{TTimeSeries}/\texttt{TFrequencySeries} signal-processing layer is being finalised in \texttt{libRIOPlus\_Signal}.} Continued HEP-GW collaboration, building on a legacy that dates back to the 1990s, remains a productive way to bridge particle physics, astrophysics, and computer science.

\section*{Acknowledgments}
This work was supported by MEXT, the JSPS Leading-edge Research Infrastructure
Program, JSPS Grant-in-Aid for Specially Promoted Research 26000005, JSPS
Grant-in-Aid for Scientific Research on Innovative Areas 2905 (JP17H06358,
JP17H06361, JP17H06364), the JSPS Core-to-Core Program A (Advanced Research
Networks), JSPS Grant-in-Aid for Scientific Research (S) 17H06133 and 20H05639,
JSPS Grant-in-Aid for Transformative Research Areas (A) 20A203 (JP20H05854), the
joint research program of the Institute for Cosmic Ray Research, University of
Tokyo, and JSPS Grant-in-Aid for Scientific Research (Nos.\ 23H01176, 23K25872,
23H04520 and 22K03614) and Grant-in-Aid for JSPS Fellows No.\ 22KF0329. This
material is based upon work supported by NSF's LIGO Laboratory, a major facility
fully funded by the National Science Foundation. We also thank D.~Cazarin, lead
author of the KFR library, for extensive discussions and for sharing the initial
benchmark code.

\end{document}